\documentclass[fleqn,usenatbib,usegraphicx]{mnras}
\usepackage{newtxtext,newtxmath}

\usepackage[T1]{fontenc}
\usepackage{ae,aecompl}

\usepackage{graphicx}	
\usepackage{amsmath}	

\usepackage{amsfonts}
\usepackage{mathtools}
\usepackage{cite}

\newcommand{\braket}[2]{\ensuremath{\left\langle #1 | #2 \right\rangle}}

\title[Eigenstates of Quasi-Keplerian Discs]{Eigenstates of Quasi-Keplerian Self-Gravitating Particle Discs}

\author[W. Melton and K. Batygin]{
Walker Melton,$^{1,2}$\thanks{E-mail: wmelton@fas.harvard.edu}
Konstantin Batygin,$^{3}$
\\
$^{1}$Deparment of Physics, Harvard University, 17 Oxford St., Cambridge, MA 02138\\
$^{2}$Division of Physics, Mathematics, and Astronomy, California Institute of Technology, 1200 E. California Blvd., Pasadena, CA 91125, USA\\
$^{3}$Division of Geological and Planetary Sciences, California Institute of Technology, 1200 E. California Blvd., Pasadena, CA 91125, USA
}

\date{Feb. 2021}

\pubyear{2021}

\begin{document}
\label{firstpage}
\pagerange{\pageref{firstpage}--\pageref{lastpage}}
\maketitle

\begin{abstract}
Although quasi-Keplerian discs are among the most common astrophysical structures, computation of secular angular momentum transport within them routinely presents a considerable practical challenge. In this work, we investigate the secular small-inclination dynamics of a razor-thin particle disc as the continuum limit of a discrete Lagrange-Laplace secular perturbative theory and explore the analogy between the ensuing secular evolution -- including non-local couplings of self-gravitating discs -- and quantum mechanics. We find the `quantum' Hamiltonian that describes the time evolution of the system and demonstrate the existence of a conserved inner product. The lowest-frequency normal modes are numerically approximated by performing a Wick rotation on the equations of motion.  These modes are used to quantify the accuracy of a much simpler local-coupling model, revealing that it predicts the shape of the normal modes to a high degree of accuracy, especially in narrow annuli, even though it fails to predict their eigenfrequencies.
\end{abstract}




\section{Introduction}



Astrophysical discs are ubiquitous, and their long-term evolution under self-gravity constitutes one of the classic problems of dynamical astronomy.   Having a generic formation channel characterized by energy dissipation at constant angular momentum \citep{GOLD:1982}, astrophysical discs exhibit significant diversity. Some discs, including active galactic nuclei and young protoplanetary discs, are primarily composed of gaseous hydrogen and helium.  In these fluid discs, self-gravity, viscosity, and pressure support all play significant roles. Other systems, including debris discs and discs of stars around supermassive black holes, have no appreciable internal pressure and undergo purely gravity-dominated evolution. 
  Moreover, when the mass of the system is dominated by the central body, the trajectories of the individual particles resemble the closed elliptical orbits of a binary system, rendering the dynamics quasi-Keplerian. 

Although astrophysical discs are often envisioned to be planar and perfectly circular, observations suggest that in reality, they can possess considerable large-scale structure. One common feature present in discs spanning a broad range of scales is warping - a process routinely attributed to an interplay between external perturbations and internal angular momentum transfer \citep{BINNEYTREMAINE}. For example, the origin of the warped stellar disc at the center of the Milky Way has been ascribed to a mix of self-gravitational effects and torques from nearby star clusters \citep{KOCSIS:2011}.  Similarly, large-scale warps of young circumstellar discs are often imagined to arise due to both self-interactions within the disc and external perturbations \citep{NESVOLD:ETAL:2016}. Gravitational torqueing of protoplanetary discs may even be responsible for planetary spin-orbit misalignments \citep{BATE:ETAL:2010, SPALDING:2014}.

In these and other quasi-Keplerian discs, self-gravitational coupling leads to long-period variations in the orbital elements.  Although qualitatively simple, obtaining quantitative descriptions of the long-term consequences of this process can present a considerable practical challenge.  In this work, we attack one aspect of this problem and consider angular momentum transport on  timescales much longer than the orbital period but considerably shorter than the lifespan of the system.  In particular, we focus on the secular inclination evolution of a dynamically cold particle disc. 

Adopting an orbit-averaged framework, we investigate the secular evolution as the continuum limit of a discrete model of 
disc dynamics, assuming that the degree of warping is small.  To this end, \citet{BAT:2018} has recently demonstrated that the dynamics captured by the Lagrange-Laplace perturbative treatment of disc evolution mirror the structure of the free-particle Schr{\"o}dinger equation when coupling between distant parts of the disc is neglected. In this local-coupling model, the dynamics are equivalent to a particle-in-a-box with Neumann boundary conditions; in terms of this analogy, propagating waves in the disc formed from a superposition of normal modes would correspond with nodal bending waves \citep{BINNEYTREMAINE}.

The local-coupling model presents a compellingly simple picture of the secular small-inclination evolution of a self-gravitating razor thin particle disc, but more work is needed to understand its limits of applicability.  Notably, the secular eigen-frequencies predicted by the naive local-coupling model are significantly lower than those seen in simulations including non-local coupling \citep{BAT:2018}.  Additionally, because asymmetric gravitational couplings between narrow rings in the disc are neglected, the local-coupling model does not in general conserve angular momentum.  Despite these shortcomings, numerical simulations indicate that they are close to the true normal modes for thin annuli, such as those that may occur during pebble accretion in protoplanetary discs \citep{MORBI:2020}. Even for wider discs, the equations of motion are almost diagonal in the local-coupling normal mode basis.

In this paper, we extend the local-coupling Schrödinger model to include non-local couplings and show that a conserved inner product exists on the space of solutions to the equations of motion. By exploiting this, we can extract close approximations of the true small-inclination normal modes.  We compare these normal modes to those predicted by the local-coupling model to understand its limits of applicability, finding that the local-coupling normal modes provide an excellent approximation to the true normal modes in narrow discs. The remainder of the paper is structured as follows.  We first introduce the secular theory of a discretized self-gravitating disc and take the limit as the discretization becomes infinitely fine to obtain an effective model for the long-wavelength excitations of a continuous disc in section 2. We then derive a conserved inner product on the space of
 solutions of the equations of motion.  The first four nontrivial lowest-frequency, or low-lying, normal modes of the disc are extracted by formally rotating time in the complex plane and these modes are compared to the local-coupling normal modes and the conventional Lagrange-Laplace normal modes found by diagonalizing the equations of motion. Our results are summarized in section 4. 

\section{Lagrange-Laplace Secular Theory: Continuum Limit} 

To understand secular angular momentum transport in razor-thin particle discs, we adopt an orbit-averaged Lagrange-Laplace perturbative treatment.  We partition the disc into a finite set of rings, interacting through a lowest-order expansion of the secular perturbing function (see chapters 6 and 7 of \citealt{MURRAYDERMOTT}), and take the continuum limit by letting the spacing of the rings approach zero while holding fixed the extent of the disc.  We then derive the equations of motion in the continuum limit and find a conserved inner product on the set of 
 solutions. 
\subsection{Parameters of the Model}  

We assume that the mass of the central body dominates the short-term dynamics of the system so that the orbits of individual particles appear Keplerian over an orbital period, with a mean motion that is well-approximated by Kepler's third law

\begin{align}
n = \sqrt{\frac{\mathcal{G}M}{a^3}},
\end{align}
 where $M$ is the mass of the central body and $a$ is the semimajor axis.  Additionally, we assume that orbital eccentricities and relative inclinations are small throughout the disc. This ensures that the velocity dispersion -- which sets the vertical thickness of the system -- is small compared to the Keplerian velocity at a fixed semi-major axis $a$. Accordingly, we consider the aspect ratio of the disc $\beta = h/a$ as a small parameter and assume that it is constant throughout the disc. Finally, we assume that the disc extends from $a_{\rm{in}}$ to $a_{\rm{out}}$, and define $\mathcal{L} = \log a_{\rm{out}}/a_{\rm{in}}$. Realistic systems have $\mathcal{L}$ on the order of a few; $\mathcal{L}$ exceeds 10 only in exceptional circumstances \citep{BAT:2018}.  Location within the disc is parametrized by $\rho = \log a/a_0$, where $a_0$ is taken to be $a_{\rm{in}}$ by convention.  The inclination and longitude of ascending node are parametrized by a complex coordinate $\eta$ such that 
\begin{equation}
\eta = \frac{i\mathrm{exp}(\iota \Omega)}{\sqrt{2}},
\end{equation}
 where $i$ is the inclination and $\Omega$ is the longitude of ascending node. $\eta_j$ refers to the complex number parametrizing the inclination and longitude of ascending node of the $j$-th ring of the discretized disc, $m_j$ the mass , $n_j$ the mean motion, and $a_j$ the semimajor axis. After taking the continuum limit, we parametrize position within the ring by the variable $\rho = \log a/a_{\mathrm{in}}$, so that $\rho$ extends from $\rho = 0$ to $\rho = \mathcal{L}$.

In our analysis of the local-coupling model, we assume that the surface density profile is described by the power law:
\begin{align}
\Sigma(a) = \Sigma_0(a_{\rm{in}}/a)^{1/2}.
\end{align}
We focus on this case to make contact with the numerical work performed in \citet{BAT:2018}, but the formalism described in this paper applies more generally, and, when applicable, we will describe the generalization to arbitrary surface density profile. 

Moreover, by virtue of being orbit-averaged, we assume that the system is gravitationally stable, implying that $m_{\mathrm{disc}} \lessapprox \beta M/2$ \citep{ARMITAGE}.

\subsection{Dynamics} 

To correct the local-coupling model and understand its successes and deficiencies, a similar method must be developed taking non-local coupling into account.  Accordingly, we will partition the disc into a finite set of rings, indexed by $i = 1, \ldots, N$, and will eventually take $N \to \infty$. In the ring-partitioned disc system, the disturbing function including all couplings is \citep{BAT:2018}
\begin{equation}
\begin{split}
\mathcal{R}_j &= B_{jj}\eta_j^*\eta_j + \sum_{i \ne j}B_{ji}(\eta_j\eta_i^* + \eta_j^*\eta_i) \\
 B_{jj} &= -\sum_{i \ne j} B_{ji} \\
 B_{jk} &= \frac{n_jm_k}{4M}\begin{cases}
\frac{a_k}{a_j}\tilde{b}_{3/2}^{(1)}(a_k/a_j) & k  \le j \\
	\frac{a_j^2}{a_k^2}\tilde{b}_{3/2}^{(1)}(a_j/a_k) & k > j
 \end{cases},
\end{split}
\end{equation}
where $\tilde{b}_{3/2}^{(1)}(\alpha)$ is a Laplace coefficient \citep{BINNEYTREMAINE}. To prevent $\tilde{b}^{(1)}_{3/2}(\alpha)$ from diverging as $\alpha \to 1$, we soften\footnote{This is a standard technique to account for the finite vertical extent of the disc; if we were performing an area integral $d\rho dz$, the result would be finite.  When we pass to a description ignoring the $\hat{z}$ axis, we must include this softening parameter to avoid unphysical singularities.} the Laplace coefficient by the disc aspect ratio $\beta$ \citep{HAHN:2003}:
\begin{equation}
\tilde{b}^{(j)}_{k}(\alpha) = \frac{2}{\pi}\int_0^\pi \frac{\cos j\psi}{(1-2\alpha\cos\psi+\alpha^2+\beta^2)^{k}}.
\end{equation} 
Because $B_{jk} \propto m_k$ for $j \ne k$, we can take the continuum limit of the disturbing function to obtain 
\begin{equation}
\begin{split}
\mathcal{R}(\rho) &= \int C(\rho,\rho')\left[\eta(\rho')^*\eta(\rho)+\eta(\rho')\eta(\rho)^*\right]dm' \\
&- \left(\int C(\rho,\rho')dm'\right)\eta(\rho)^*\eta(\rho) \\
C(\rho,\rho') &= \frac{n(\rho)}{4 M}\begin{cases}
e^{\rho'-\rho}\tilde{b}^{(1)}_{3/2}(e^{\rho'-\rho}) & \rho' \le \rho \\
e^{2\rho-2\rho'}\tilde{b}^{(1)}_{3/2}(e^{\rho-\rho'}) & \rho' > \rho
\end{cases}.
\end{split}
\end{equation}

 Applying the equation of motion $\iota\frac{\partial\eta}{\partial t} = -\partial\mathcal{R}(\rho)/\partial \eta(\rho)^*$, we find that 
\begin{equation}
\begin{split}
\iota\frac{\partial\eta(\rho)}{\partial t} &= \int C(\rho,\rho')(\eta(\rho)-\eta(\rho'))dm' \\
&= \int 2\pi a_{in}^2e^{2\rho'}\Sigma(a_{in}e^{\rho'})C(\rho,\rho')(\eta(\rho)-\eta(\rho'))d\rho' \\
&= \int B(\rho,\rho')(\eta(\rho)-\eta(\rho'))d\rho'.
\end{split}
\end{equation}
This can be written in the form of a general time-dependent Schr{\"o}dinger equation with Hamiltonian $\mathcal{H}$, in which 
\begin{equation}
\mathcal{H}\eta = \int B(\rho,\rho')(\eta(\rho)-\eta(\rho'))d\rho'.
\end{equation}
If we expand $\eta(\rho')$ in a power series around $\rho$, we find that 
\begin{equation}
\begin{split}
\mathcal{H} &= -\sum_{k=1}^\infty c_k(\rho)\frac{\partial^k}{\partial\rho^k} \\
c_k(\rho) &= \frac{1}{k!}\int B(\rho,\rho')(\rho'-\rho)^kd\rho'.
\end{split}
\end{equation}
We should stress that we have so far only assumed that the relative inclination is perturbatively small throughout the disc and that the relevant dynamics are secular.  We have not assumed anything about the coupling between different parts of the disc; hence, the equation of motion presented above is equivalent to the Lagrange-Laplace equations of motion written in terms of the complex variable $\eta$. 

We obtain the local-coupling equation by keeping only the term proportional to $\partial^2/\partial\rho^2$ and averaging over the spatial dependence of $c_2(\rho)$:
\begin{equation}
\begin{split}
\iota\frac{\partial\eta}{\partial t} &= \mathcal{H}_{LC}\eta = -c_2\frac{\partial^2\eta}{\partial \rho^2} \\
c_2 &= \frac{1}{\mathcal{L}}\int_0^\mathcal{L} c_2(\rho)d\rho.
\end{split}
\end{equation}

In the interior of the disc, $c_k$ is small for odd $k$ due to the approximate symmetric coupling, while $c_k$ remains significant for even $k$, so that the dominant contribution to the equations of motion has $k = 2$.  While $c_k$ tends to drop in magnitude as $k$ increases, it is worth noting that the higher-order contributions can become significant for states with large wavenumbers.  	

\subsection{The Conserved Inner Product} 

To extend the analogy between the small-inclination dynamics of a self-gravitating particle disc and the structure of quantum mechanics, we need not only the emergence of a general time-dependent Schr{\"o}dinger equation but also the presence of a conserved inner product.  Many of the tools of quantum mechanics rely on the existence of a set of orthogonal eigenmodes with real oscillation frequencies, which is guaranteed by the conservation of some inner product $\braket{\psi}{\phi}$.  This will be true if the Hamiltonian is Hermitian with respect to the given inner product. Correspondingly, in this section we first show that there exists an inner product such that $\braket{\eta}{\eta}$ is conserved by requiring that angular momentum is conserved; then, we show that $\braket{\eta}{\nu}$ must be conserved as well for any solutions to the equations of motion $\eta$ and $\nu$.  

Consider the angular momentum of a small ring of mass $dm$ and semi-major axis $a$.  Its $\hat{z}$ angular momentum deficit is, in the small inclination approximation, $\sqrt{GMa}i^2dm/2$ \citep{LASKAR:1997}.  Integrating over all infinitesimal rings, the total $z$-angular momentum deficit is 
\begin{equation}
\int \frac{\sqrt{GMa}i^2dm}{2} \propto \int 2\pi a\Sigma(a)\sqrt{a}i^2da \propto \int a\Sigma(a)\sqrt{a}i^2da = \braket{\eta}{\eta}.
\end{equation}
For the choice of disc parameters above, $\Sigma(a)\sqrt{a}$ is constant, and the inner product is given by (dropping an insignificant constant)
\begin{equation}
\braket{\eta}{\eta} = \int a\eta^*\eta da.
\end{equation}

For these dynamics to mirror completely the structure of quantum mechanics, it must be that $\braket{\eta}{\nu}$ is conserved for any two solutions $\eta$ and $\nu$.  Indeed, this is the case for stable dynamics such as those described by the Lagrange-Laplace model. Because the `matrix elements' $B(\rho, \rho')$ of $\mathcal{H}$ are real,  $(\mathcal{H}\eta)^* = \mathcal{H}[\eta^*]$.  Thus, if $\eta$ is a normal mode with frequency $\omega$, $\eta^*$ must be a normal mode with frequency $\omega^*$.  For stable dynamics, $\omega = \omega^*$ so that the amplitude of normal modes does not grow with time..  Hence, $\eta^* + \eta$ is also a normal mode, and the normal modes can be taken to be real functions of the coordinate $\rho$.  Suppose that $\eta_a$ and $\eta_b$ are two real and distinct eigenfunctions.  Then, $\braket{\eta_a + \eta_b}{\eta_a+\eta_b}$ is conserved, which, by the conservation of $\braket{\eta}{\eta}$, implies that $\braket{\eta_a}{\eta_b} = -\braket{\eta_b}{\eta_a} = -\braket{\eta_a}{\eta_b}^*$.  Therefore, $\braket{\eta_a}{\eta_b} = \gamma \iota$ for $\gamma \in \mathbb{R}$.  $\braket{\eta_a}{\eta_b}$ must also be real, however, so $\gamma\iota \in \mathbb{R}$.  The only way both conditions can be satisfied is if $\braket{\eta_a}{\eta_b} = 0$ for unequal eigenstates.  Now, any solution can be written as the sum of eigenstates $\eta = \sum_a \alpha_a\eta_a$, $\nu = \sum_a \beta_a\eta_a$.  Then, 
\begin{equation}
\begin{split}
\braket{\eta}{\nu} &= \sum_{a,b}\alpha^*_a\beta_b\braket{\eta_a}{\eta_b} \\
&= \sum_a \alpha^*_a\beta_a\braket{\eta_a}{\eta_a}.
\end{split}
\end{equation}
Because this quantity is independent of time, the inner product is conserved, provided that the model conserves angular momentum.  We will provide a direct proof of the conservation of this inner product in Appendix \ref{DIP}.

For a general surface density profile $\Sigma(a)$, it is easy to extend the analysis above to show that the inner product
\begin{equation}
\label{IPE}
\braket{\eta}{\nu} \propto \int \eta^*(a)\nu(a)a^{3/2}\Sigma(a)da
\end{equation}
is conserved. This completes the analogy with quantum mechanics: not only do the dynamical equations take the form of a generalized time-dependent Schrödinger equation, the space of 
 solutions of the equations of motion possesses a conserved inner product. 	This implies that the Hamiltonian is Hermitian with respect to the inner product above.

\begin{figure}
\begin{center}
\includegraphics[width=\columnwidth]{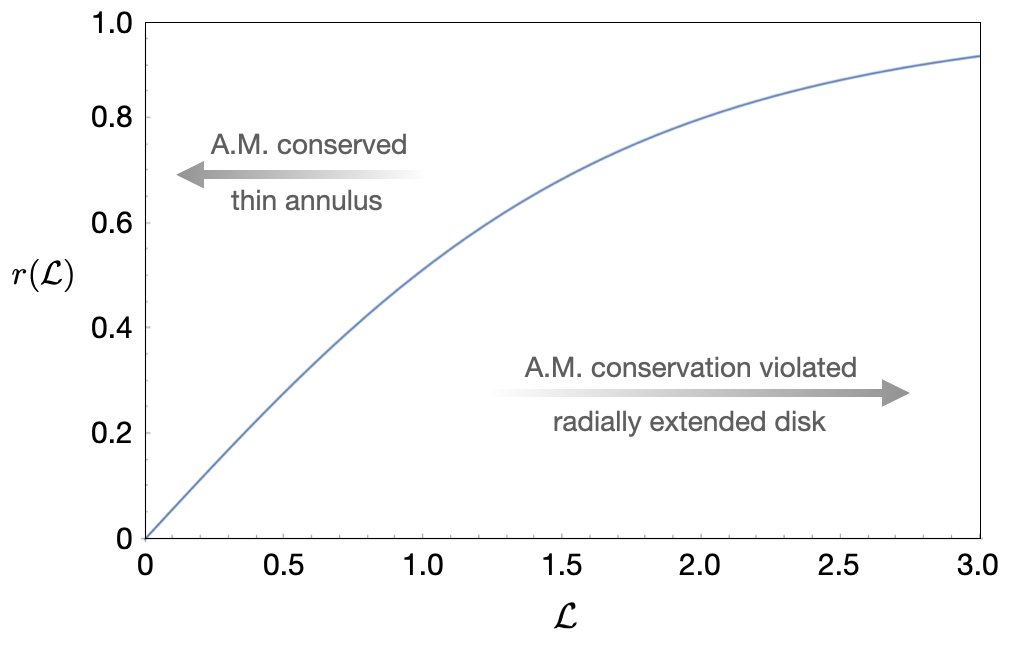}
\end{center}
\caption{A quantitative measure of the non-conservation of angular momentum wihtin the framework of the local coupling model.  The quantity $r(\mathcal{L})$ is shown for a disc with an inverse square root density profile from $\mathcal{L} = \log a_{\mathrm{out}}/a_{\mathrm{in}} = 0$ to $\mathcal{L} = 3$. Note that the degree of angular momentum non-conservation is only small for $\mathcal{L} \ll 1$, implying that the local model is applicable to annuli but not extended discs.}
\label{fig:rLfig}
\end{figure}

It is worth noting that Equation \ref{IPE} is not the same inner product as that preserved by the local coupling model.  The local coupling model preserves the inner product 
\begin{align*}
\braket{\eta}{\nu} \propto \int_0^{\mathcal{L}} \eta^*(\rho)\nu(\rho)d\rho
\end{align*}
which is the sum of the angular momentum deficit of each infinitesimal ring with a nontrivial weighting.  The source of this difference is the assumption that the coupling was symmetric when deriving the local coupling model.  As the true normal modes of the disc are orthogonal with respect to the inner product given by Equation \ref{IPE} rather than that preserved by the local coupling model, this inner product is not preserved by the true, non-local dynamics. As a benchmark of the failure of the local coupling model to conserve angular momentum, we compute a measure of the variation of angular momentum if the local coupling model was correct.  Since we focus on the long-wavelength normal modes of the disc, we examine the failure of angular momentum to be conserved in states dominated by the longest-wavelength normal modes.  If the local coupling model was correct, we could consider a system with state $\eta(\rho,t) = c_0(\rho)e^{-\iota\omega_0t} + c_1(\rho)e^{-\iota\omega_1t}$, letting $c_\ell(\rho)$ be the local coupling modes.  Then, to linear order, the angular momentum of this state is $\braket{\eta}{\eta} = \braket{c_0}{c_0} + \braket{c_1}{c_1} + \braket{c_1}{c_0}e^{\iota(\omega_1-\omega_0)t} + \braket{c_0}{c_1}e^{\iota(\omega_0-\omega_1)t} = \braket{c_0}{c_0} + \braket{c_1}{c_1} + 2\braket{c_0}{c_1}\cos(\omega_0-\omega_1)t$, so that the width of the variation of angular momentum is $2\braket{c_0}{c_1}$. We thus define
\begin{equation}
r(\mathcal{L}) = \frac{|\braket{c_0}{c_1}|}{\sqrt{\braket{c_0}{c_0}\braket{c_1}{c_1}}}
\end{equation}
which provides a measure of the extent of variation around the average angular momentum that is insensitive to the normalization of the inner product and the normalization of the solutions $c_0, c_1$. For a disc with an inverse square root surface density profile,
\begin{equation}
r(\mathcal{L}) = \frac{4\mathcal{L}^2\sqrt{2-2\mathcal{L}^2/(\pi^2+2\mathcal{L}^2)}\coth\mathcal{L}}{\pi^2+4\mathcal{L}^2}
\end{equation}
which is plotted in Figure (\ref{fig:rLfig}).
The error is small (on the order of 5\%) for $\mathcal{L} \lessapprox 0.1$, so the local coupling model may be expected to hold for narrow annuli, but becomes progressively inaccurate as $\mathcal{L}$ increases.s

\section{Normal Modes of the disc} 

The non-local equations of motion found in the previous section are not amenable to analytical techniques in the same way as the local model.  Variational methods can be applied, but without a clear understanding of the space of states described by these dynamics, we risk finding false solutions.  Additionally, these results may be sensitive to the choice of family of trial functions.  An approach that extracts the normal modes without requiring assumptions about their form would be far more useful.

Such a method can be found by rotating time in the complex plane, turning the oscillatory solutions of the true equations of motion into exponentially damped solutions.  By applying the imaginary time equations of motion and the orthogonality rules derived above, we can successively extract the low-lying normal modes of the disc.  In this section, the imaginary time formalism is described and extended to the equations of motion for a disc composed of a large but finite number of rings.  

\subsection{The Imaginary Time Formalism} 

In real time, the equations of motion are 

\begin{equation}
\begin{split}
\iota\frac{\partial\eta}{\partial\tau} &= \mathcal{H}\eta \\
\eta(\rho,t) &= e^{-\iota\mathcal{H}t}\eta(\rho,0).
\end{split}
\end{equation}
The solutions take the familiar form $\eta(\rho,t) = \sum_k c_k\eta_k(\rho)e^{-\iota\omega_kt}$. When we perform a Wick rotation, sending $t \to -\iota\tau$, the equations of motion take the form 
\begin{equation}
\begin{split}
\frac{\partial\eta}{\partial\tau} &= -\mathcal{H}\eta \\
\eta(\rho,\tau) &= e^{-\mathcal{H}\tau}\eta(\rho,0).
\end{split}
\end{equation}

These equations have solutions of the form $\eta(\rho,\tau) = \sum_k c_k\eta_k(\rho)e^{-\omega_k\tau}$ where $\eta_k(\rho)$ is the shape of the $k$-th normal mode.  These solutions are exponentially damped, with higher-frequency normal modes decaying more quickly than lower-frequency modes.  At large $\tau$, $\eta(\rho,\tau)$ will be dominated by the normal mode of lowest frequency whose coefficient in the mode expansion above is non-zero.

By choosing $\eta(\rho,0)$ such that $\braket{\eta(0)}{\eta_k} = 0$ for $k < \ell$ and $\braket{\eta(0)}{\eta_\ell} \ne 0$, $\eta(\rho,\tau)$ will be dominated by $\eta_\ell(\rho)$ for large $\tau$.  Because the ground state $\eta_0(\rho) = 1$ is known by symmetry, we can obtain low-lying normal modes in turn by applying successive Gram-Schmidt orthogonalization, which allows us to remove contributions from lower-frequency normal modes. Even though continuum descriptions of the disc are appealing, to apply this formalism numerically, the Schr{\"o}dinger-evolution model derived above must be cast into a discrete form.

While the aforementioned non-local model applies to continuum discs, a conceptually similar model governs the evolution of a discrete ring-partitioned disc.  
Rather than being described by a function $\eta(\rho)$, the state of the disc will be described by a finite set of complex numbers $\eta_j$, with $j = 1 \ldots N$.  The Hamiltonian is a linear function of the space of states with coefficients given by the discrete counterpart of equation 4.3, and the inner product is 
\begin{equation}
\braket{\eta}{\nu} = \sum_j m_j\sqrt{a_j}\eta^*_j\nu_j
\end{equation} 
so that $\braket{\eta}{\nu}$ is proportional to the $\hat{z}$ angular momentum deficit. We will use the equations of motion for this discretized model to extract the normal modes of the disc. 

\subsection{Numerical Simulations: the First Four Non-Zero Normal Modes of a $\Sigma \propto a^{-1/2}$ disc} 
A distinct prediction made by the local coupling model outlined in \citet{BAT:2018} and recalled in section 2.2 is that the normal modes of the disc have the form:
\begin{equation}
\eta_k(\rho) = \cos\frac{k\pi\rho}{\mathcal{L}} = \cos\left(\frac{k\pi}{\mathcal{L}}\log\frac{a}{a_0}\right).\label{eq:lcnm}
\end{equation}
Correspondingly, in order to quantify the accuracy of the local-coupling model, we simulated a trial disc numerically. For definitiveness, we adopted the same example as that considered by \citet{BAT:2018}. The disc had total mass $1 M_\oplus$, orbited a star of $1\ M_\odot$, and extended from $a = 1\ \mathrm{AU}$ to $a = 1.1\ \mathrm{AU}$. Notably, this system is similar to the annulus of solid debris, within which the formation of the solar system's terestrial planets unfolds \citep{HANSEN:2009}. The surface density profile was taken to be $\Sigma(a) \propto a^{-1/2}$ as in section 2.1. The aspect ratio of the disc was $\beta = h/a = 10^{-3}$. In the simulation, the disc was partitioned into a series of $N = 100$ rings, each of which coupled gravitationally to every other ring.  

\begin{figure*}
\begin{center}
\includegraphics[width=\textwidth]{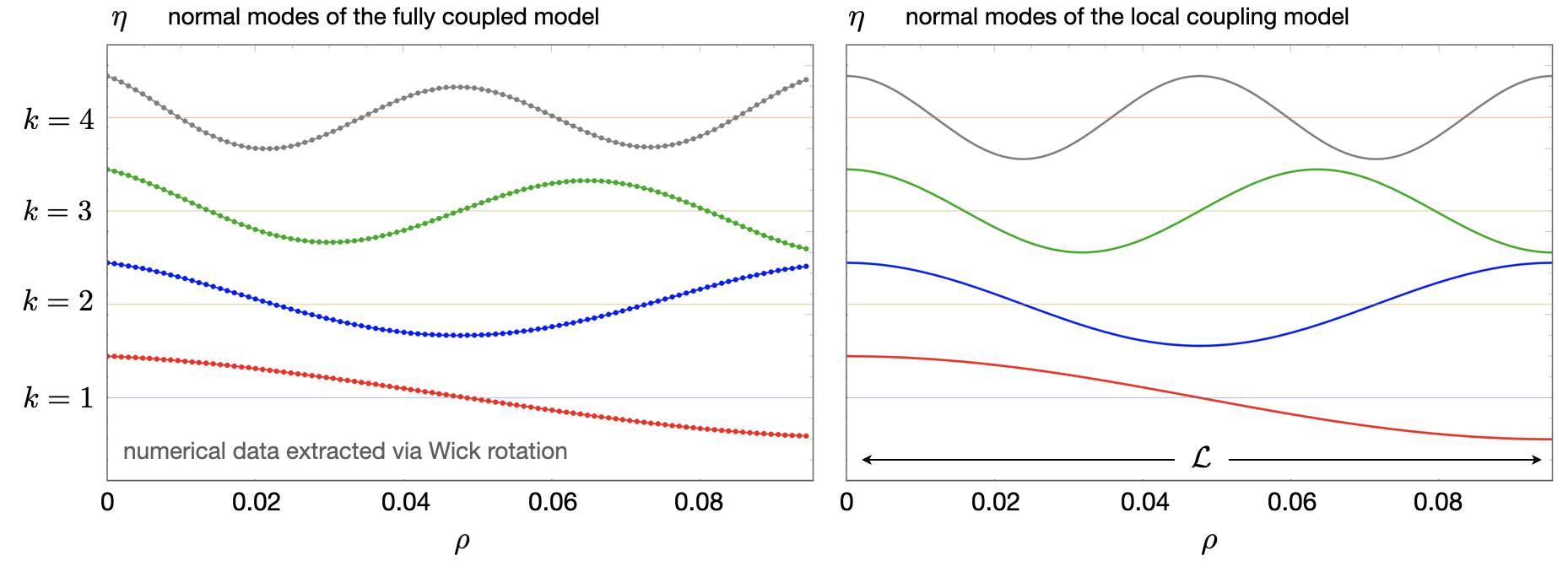}
\caption{The first four non-constant normal modes of the local coupling model (right) and the corresponding numerically extracted modes for the fully coupled model with the same disc parameters (left).  The red line is the lowest-frequency non-constant mode ($k = 1$ in the local coupling model), blue the second lowest frequency mode ($k = 2$), green the third lowest ($k = 3$), and gray the fourth lowest ($k = 4$). For each model, the simulated disc spans a semi-major axis range of 1-1.1 AU and is comprised of $N=100$ equally spaced, interacting wires. The aspect ratio is taken to be $\beta=0.001$ throughout. The cumulative disc mass is equal to that of the Earth and the mass of the central body is equal to that of the sun.}
\label{fig:plotfig}
\end{center}
\end{figure*}

The imaginary time equations of motion $\partial\eta/\partial \tau = -\mathcal{H}\eta$ were solved using a conventional 4th-order Runge-Kutta integration.  The initial conditions were chosen so that $\eta(\rho,0)$ was close to $\cos k\pi \rho/\mathcal{L}$ for $k = 1,2,3,4$, with contributions from the lower-frequency normal modes removed using Gram-Schmidt orthogonalization. To prevent the desired state from becoming damped below machine precision, the Hamiltonian was shifted by a term proportional to the identity so that the frequency of the desired mode was close to 0.  The calculations were run for several thousand years of simulation time, by which point higher-frequency modes had damped to the point that they were negligible compared to the mode of interest. After the state was extracted from the simulation, Gram-Schmidt orthogonalization was applied again to remove residual contributions from low-lying modes that grew exponentially.  The modes derived were then used as the input for a real-time simulation; upon completing this exercise, no significant evolution of the magnitude of the inclination was observed, indicating that these modes are indeed very close to the true normal modes of the system. The results of these simulations are presented in Figure \ref{fig:plotfig}.

The periods of these normal modes were then estimated from the real time simulations; the results are presented in Table \ref{tab:freq1} along with best fit wavenumbers $k$. We reiterate that these normal modes were not derived by assuming a specific form of the normal modes beyond the initial condition $\eta(\tau = 0) = \cos k\pi \rho/\mathcal{L}$, and the results are generally insensitive to the initial choice as long as the correct orthogonalization procedures are followed.  Hence, we use these normal modes to test the accuracy of the local-coupling normal mode approximations, which given by Equation \ref{eq:lcnm}. 
\begin{table}
\begin{center}
\caption{The periods and best-fit wavenumbers of the first numerically extracted modes. \label{tab:freq1}}
\begin{tabular}{@{}lcc@{}}
\textbf{i} & $k$ & $T_i = \frac{2\pi}{\omega_i}$ (years) \\ \hline
1 & 24.1748 & 3947 \\ 
2 & 61.2425 & 1670 \\
3 & 90.6545 & 1041\\
4 & 120.424 & 756
\end{tabular}
\end{center}
\end{table}

While this may seem a numerically tedious way of extracting the normal modes, convergence to the true normal mode is fast as the difference to the true normal modes falls exponentially in time.  As such, when it is computationally simpler to simulate a system over a small number of periods than to numerically diagonalize the Lagrange-Laplace equations of motion (as becomes progressively the case with increasing $N$), this will provide a useful way of numerically extracting low-lying modes of the disc. 

\subsection{Comparison with Traditional Lagrange-Laplace Normal Modes} 

Now that we have used the Wick-rotated equations of motion to numerically evaluate the normal modes, we compare to the results found by diagonalizing the Lagrange-Laplace equations of motion.  Specifically, we will show that the normal modes for the $N = 100$ disc simulated here are close to the modes found by explicitly diagonalizing the equations of motion.

The Lagrange-Laplace equations of motion take the form 
\begin{equation}
\frac{d\eta_j}{dt} = -\iota B_{jk}\eta_k
\end{equation}
$B_{jk}$ was explicitly diagonalized and its eigenvalues and eigenvectors explicitly computed.  The periods predicted by explicit diagonalization of $A_{ij}$ are compared to those predicted from the imaginary-time simulations in Table \ref{tab:freqcomp}.  The agreement between the predictions derived using the imaginary time formalism are clearly excellent approximations to those derived by explicit diagonalization of the normal modes. 

\begin{table}
\begin{center}
\caption{The periods in years from explicit diagonalization and the imaginary time simulation \label{tab:freqcomp}}
\begin{tabular}{@{}lcc@{}}
\textbf{i} & $T_i$ (Imaginary Time) & $T_i$ (Explicit Diagonalization) \\ \hline
1 & 3947 & 3947 \\
2 & 1670 & 1671 \\
3 & 1041 & 1041 \\
4 & 756 & 757
\end{tabular}
\end{center}
\end{table}

The normal modes derived from the imaginary time simulation and from diagionalizing the equations of motion, normalized so that $\eta(0) = 1$, are shown in Figure \ref{fig:diagcomp} below. Differences between the exact modes of the discrete system and those derived from the imaginary time procedure were on the order of $10^{-10}$ to $10^{-3}$ as a fraction of the exact mode, with particularly strong agreement for the two lowest normal modes. Better agreement could be found by running the imaginary time simulation for a longer period of time. We conclude that the modes derived using the imaginary time simulations are in excellent agreement with the modes found by explicitly diagonalizing the equations of motion, validating our method, that the Gram-Schmidt orthogonalization is sufficiently reliable, and indicating that we can use them to quantify the accuracy of the local coupling modes.
\begin{figure*}
\centering
\includegraphics[width=\textwidth]{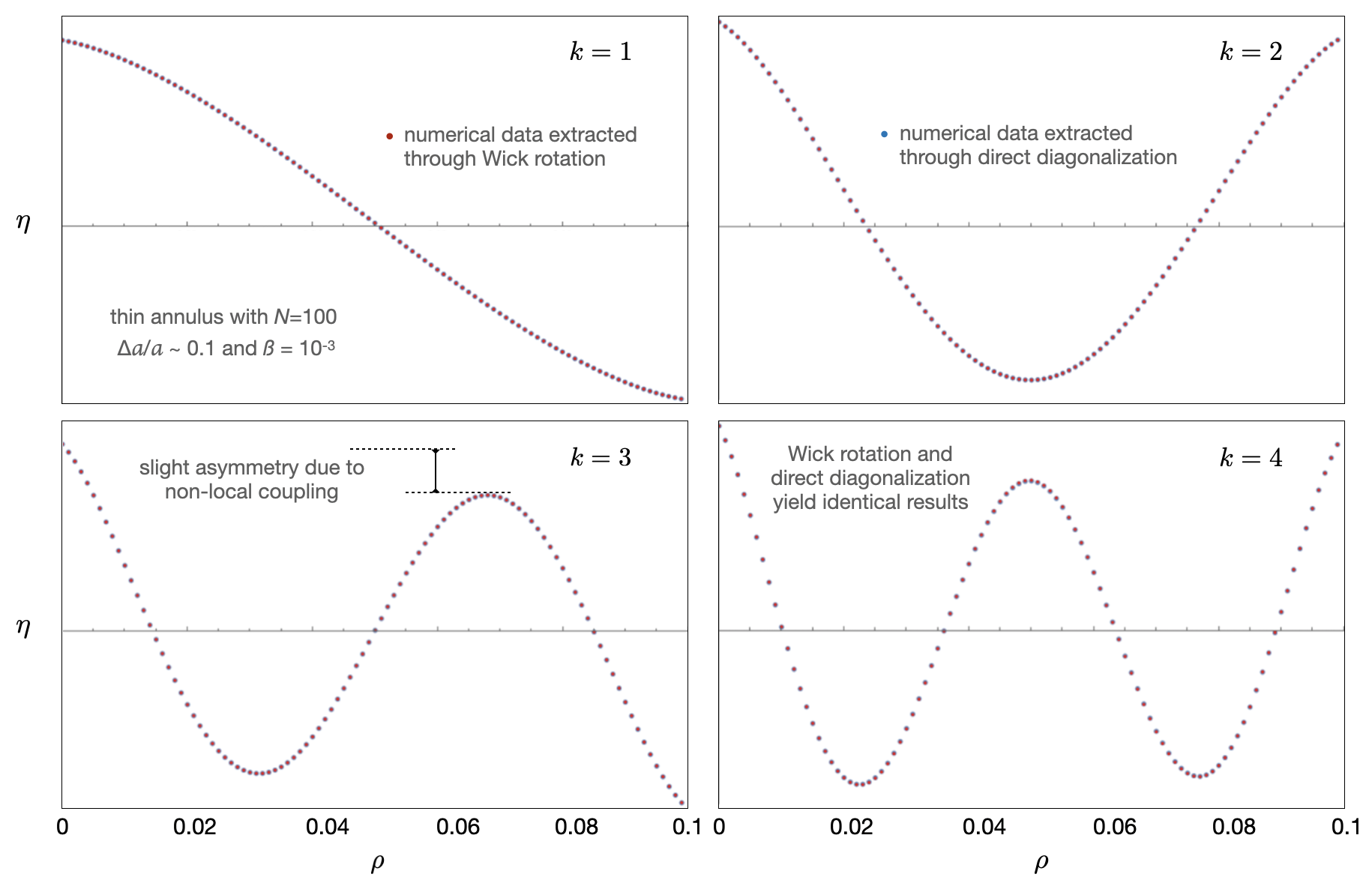}
\caption{The first four normal modes as extraced by the imaginary time formalism (red) and direct diagonalization (blue).  The two are in excellent agreement.} 
\label{fig:diagcomp} 
\end{figure*}

\subsection{The Accuracy of the Local-Coupling Normal Modes} 
Although the curves delineated in Figure \ref{fig:plotfig} appear close to the prediction of the local model (Eq. \ref{eq:lcnm}), their correspondence is not exact.  But just how inexact are they?	 Having strong estimates of the true normal modes of the disc, we can now numerically measure the accuracy of these local-coupling modes.

Let us consider a few different measures that are sensitive to different aspects of the local-coupling model.  The first, based on the Fourier series expansion of the true normal mode, is sensitive to both the sinusoidal nature of the mode as well as the boundary conditions on the disc.  Others, based on the accuracy of sinusoidal fits to the true normal mode, are insensitive to the boundary conditions.


\subsubsection{Power Spectrum Measures of the Accuracy of the Local-Coupling Normal Modes} 

If we assume that $\eta_\ell(\rho)$ is the true normal mode most closely approximated by $\cos \ell\pi \rho/\mathcal{L}$, we can expand $\eta_\ell(\rho)$ as a Fourier series: 
\begin{equation}
\eta_\ell(\rho) = \eta_0 + \sum_k c_k\cos k\pi \rho/\mathcal{L} + \sum_q s_q\sin q\pi \rho/\mathcal{L}.
\end{equation}
Then, if $\eta_\ell$ is well approximated by $\cos \ell\pi\rho/\mathcal{L}$, we expect $c_\ell \gg c_{\ell'}, s_k$ in which $\ell',k \in \mathbb{N}$ and $\ell' \ne \ell$.  As $\eta_\ell$ becomes more and more like the local-coupling mode, $c_\ell$ will approach unity for a properly normalized state and $c_{\ell'}, s_k$ will approach 0.  As such, the number
\begin{equation}
f_\ell = \frac{|c_\ell|^2}{\sum_k |c_k|^2 + |s_k|^2} = \frac{|c_\ell|^2}{\frac{2}{\mathcal{L}}\int d\rho |\eta_\ell(\rho)|^2}
\end{equation}
provides a reasonable measure of the similarity of the $\ell$-th true normal mode to the local-coupling normal mode.  As $\int d\rho \cos k\pi\rho/\mathcal{L} \cos q\pi\rho/\mathcal{L} = \mathcal{L}\delta_{kq}/2$,  
\begin{equation}
c_\ell = \frac{2}{\mathcal{L}}\int_0^\mathcal{L} d\rho \eta_\ell(\rho)\cos\ell\pi\rho/\mathcal{L}.
\end{equation}
Hence, 
\begin{equation}
f_\ell = \frac{2}{\mathcal{L}}\frac{\left(\int d\rho \eta_\ell(\rho)\cos \ell\pi\rho/\mathcal{L}\right)^2}{\int_0^\mathcal{L} \eta_\ell(\rho)^2}.
\end{equation}

The values of $f_\ell$ for $\ell = 1,2,3,4$ were computed for the trial disc simulated above; the results are reported in Table \ref{tab:acc}. For the low-lying normal modes in the trial disc, the local-coupling normal modes represent by far the dominant fraction of the power of the true normal modes.  Interestingly, $f_k$ appears to monotonically decrease as $k$ increases, which may indicate that the local-coupling normal mode provides a more accurate model of large-wavelength excitations than small-wavelength excitations, although our methods of extracting the normal modes become infeasible for investigating these high-frequency modes.

\subsubsection{Best-Fit Residual Measures of the Accuracy of the Local-Coupling Normal Modes} 

Because the coefficient $f_k$ represents only the contribution from cosine terms to the power spectrum, it is sensitive to assumptions about the boundary conditions on the disc; in the local-coupling model \citep{BAT:2018}, these are taken to be the Neumann boundary conditions $\partial\eta/\partial\rho = 0$ for $\rho = 0, \mathcal{L}$.  However, the local-coupling model may still provide an accurate picture of the normal modes even if the boundary conditions differ from these Neumann conditions.  It is therefore important to develop a measure of the accuracy of the local-coupling model that is sensitive only to the sinusoidal nature of the normal modes and not to the specific choice of boundary conditions on those modes. 

One measure of the accuracy of the fit is the root-mean-square of the residuals when a sinusoid is fit to a properly normalized state. To carry out this analysis, a standard least-squares algorithm was used to fit a function of the form $A\sin kx + B\cos kx$ to each of the first four numerically extracted normal modes, each normalized so that $\eta(\rho = 0) = 1$.  We should expect that the residuals are proportional to the magnitude of the state, so this is a reasonable measure of the size of deviations from harmonicity that is insensitive to the normalization of the state. This method was applied to the first four non-constant normal modes numerically extracted from the trial disc.  The results are presented in the middle column of Table \ref{tab:acc}.

Another reasonable measure of the accuracy of the fit is the adjusted $R^2$ value.  The results of these measurements are reported in the final column of Table \ref{tab:acc}.
\begin{table}
\centering
\caption{The Accuracy of the Local-Coupling Normal Modes for a Thin Annulus}
\label{tab:acc}
\begin{tabular}{c|c|c|c}
\textbf{$\ell$} & $f_\ell$ & RMS & $R^2$ \\ \hline
1 & 0.994026 & 0.0170737 & 0.99967  \\
2 & 0.985995 & 0.0911971 & 0.990557 \\
3 & 0.96852  & 0.107363 & 0.986908 \\
4 & 0.943643 & 0.126079 & 0.981944\\ 
\end{tabular}
\end{table}
Just as in the power-spectrum measure of the accuracy of the local-coupling model, these measures indicate that the accuracy of the local-coupling model decreases with increasing wavenumber. More investigation is needed for $\ell \ge 5$.

\subsection{The Variation of the Accuracy of the Local-Coupling Normal Modes with the Radial Extent of the disc} 

Solutions of the form $\sum_k c_ke^{\iota\omega t}\cos k\pi\rho/\mathcal{L}$ do not conserve angular momentum if $c_k \ne c\delta_{k,k_0}$ for some $c, k_0$.  For the trial disc simulated, the disc was rather narrow, with $a_{\rm{out}}/a_{\rm{in}} = 1.1$.  For such a narrow disc, the local-coupling modes are close to orthogonal, with only small variations of angular momentum over the course of a period.  As the disc becomes wider, however, this failure of angular momentum conservation can become more significant.  It is thus reasonable to question whether the excellent matches between the local-coupling and general-coupling modes found here are an artifact of the narrow-disc limit.

\begin{figure}
\centering
\includegraphics[width=\columnwidth]{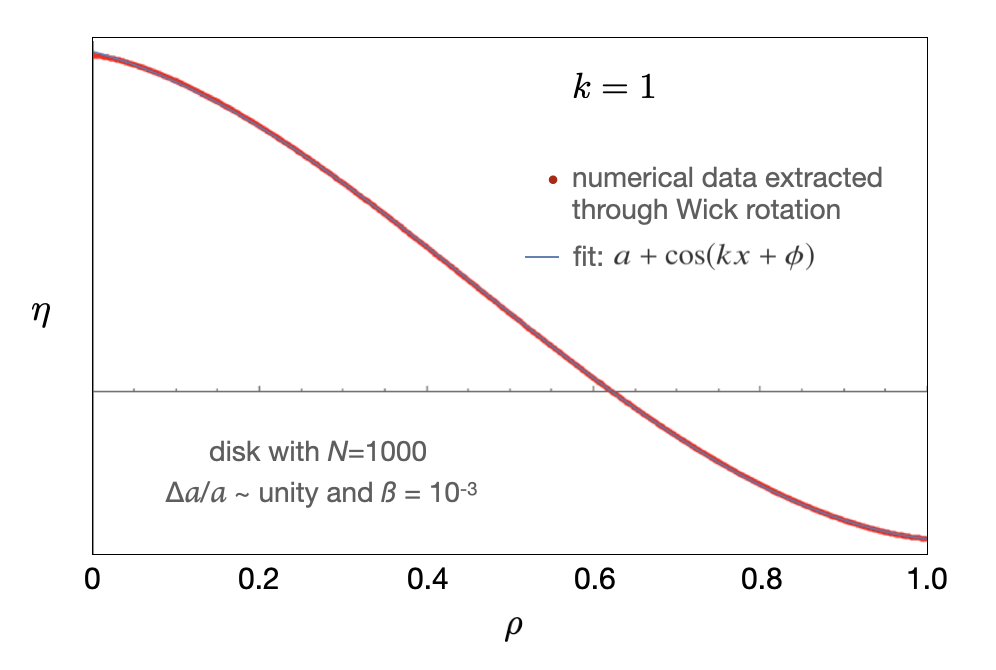}

\caption{The first normal mode of a disc with $N = 1000$ (red) and the best fit curve of the form $a + \cos (kx + \phi)$ (blue).}
\label{fig:thgsp}
\end{figure}

To investigate this possibility, the first normal modes of discs with the same density profile were simulated for different values of the width of the disc.  More specifically, this was done by keeping the separation between adjacent wires fixed, but increasing or decreasing the number of rings of the partition.  The results for $N = 50, 100, 150, 300, 1000$ are presented in Table \ref{tab:table4}.
\begin{table}
\centering
\caption{The variation of the accuracy of local-coupling modes with disc size for $k = 1$}
\label{tab:table4}
\begin{tabular}{c|c|c|c}
N & $f_1$ & RMS Residual & R$^2$ \\  \hline
50 & 0.995754 & 0.00847134 & 0.99992  \\
100 & 0.994026 & 0.0170737 & 0.99967  \\
150 & 0.990185 & 0.0246727&  0.999286\\ 
300 &0.972407 & 0.0728  & 0.99446 \\	
1000 & 0.777252 & 0.2818 & 0.993884
\end{tabular}
\end{table}
While the agreement between the local-coupling modes and the general normal modes remains high, there are indications that the local-coupling normal modes become less accurate as the width of the disc increases, particularly for $n_b \ge 1000$.  Even in this case, however, the difference between the numerical normal mode and the corresponding local mode is a constant shift (corresponding to a mixing between the lowest-energy and first non-zero mode), so even for larger discs the Hamiltonian derived is almost diagonal in the local-coupling mode basis. This is well-illustrated by the $\ell = 1$ mode of the thousand ring disc, which is to an excellent approximation a superposition of the $\ell = 0, \ell = 1$ local coupling modes for some boundary condition, as shown in Figure \ref{fig:thgsp}.  

To summarize the results in this section, the overall the agreement for the low-lying normal modes appears strong for narrow annuli, although each of the measures indicates that the agreement decreases as the wavenumber increases. Although local coupling eigenfunctions still offer adequate results for disc widths of the same order the mean semi-major axis, their degradation with increasing $\mathcal{L}$ suggests that for $\Delta a/a$ significantly in excess of unity, these expressions become highly approximate. Based upon the results summarized in Table \ref{tab:acc}, we further expect that the agreement will gradually worsen with decreasing wavelength. Despite this, the local-coupling basis remains useful as the Hamiltonian is almost diagonal in the local-coupling normal mode basis. Even this level of agreement may be surprising, however, as the local-coupling model is a drastic simplification of the notoriously rich dynamics of self-gravitating discs. 

\section{Discussion} 

The analogy between the secular small-inclination evolution of a self-gravitating razor-thin particle disc and the time-dependent Schr{\"o}dinger equation of quantum mechanics was extended beyond the local-coupling model developed in \citet{BAT:2018}.  While the time-dependent Schrodinger equation $\iota\dot{\psi} = \mathcal{H}\psi$ is most often seen in quantum mechanical regimes, it can appear more generally in situations in which the dynamics are linear and conservative.  In such a case, the map $t \mapsto U(t,t_0)$ where $U(t)$ is the generator of a unitary representation of the Lie group $\mathbb{R}$.  The Hamiltonian, then, is the generator $\mathcal{H} = \iota dU(t)/dt$ \citep{LIE}.  

In this work, we have demonstrated that the stable small-inclination dynamics of a self-gravitating disc form a unitary representation of the time-translation Lie group $\mathbb{R}$.  This structure and the deeper analogy with quantum mechanics provided by the existence of the conserved inner product $\braket{ \cdot }{ \cdot }$ give us analytical and numerical tools to investigate the stable, secular inclination dynamics of particle discs, both when the parameters of the disc are constant and when they vary on a secular timescale.

By rotating time in the complex plane, we extracted the normal modes of the disc.   This rotation turned the conservative, oscillatory dynamics of the real-time equations of motion into a purely dissipative dynamical system in which higher-frequency normal modes decay away.  While explicit diagonalization is also feasible for the discs discussed in this paper, the imaginary-time formalism provides a numerical method that may prove useful for systems where explicit diagonalization is computationally disadvantageous.

After using this formalism to find assumption-free approximations for the first four non-constant normal modes, Fourier analysis and least-squares fitting were used to estimate their similarity to the purely sinusoidal normal modes predicted by the local-coupling model.  While there were indications that the agreement between the two sets of normal modes decreases with increasing wavenumber and with the width of the disc, we found that the modes are indeed very close to the sinusoidal modes predicted by the local coupling model for narrow annuli, indicating that there is an interesting region of parameter space where the local-coupling normal modes, which have an explicit analytical form in the continuum limit, well-approximate the true normal modes of the disc. Even in wider discs, the equations of motion for the disc are almost diagonal when written in the local-coupling mode basis, which can dramatically simplify numerical calculations.

This quantum-mechanics-like formalism also provides a convenient tool for estimating the response of discs to external forces.  If the perturbing system preserves the rotational symmetry of the disc, the perturbing Hamiltonian generated by the external system will be Hermitian, and quantum perturbation theory can be used to estimate how the frequencies and shapes of the normal modes will change and provide a measure of the rigidity of the disc to external perturbations. This perturbative technique is justified by the existence of a `quantum' Hamiltonian that preserves an inner product, as has been shown in this paper.

This work clarifies the applicability of the local coupling model.  In thin, narrow discs whose dynamics are dominated by self-gravitational effects, , the local-coupling normal modes provide an excellent approximation to the true normal modes of the disc. These systems include, for example, narrow annuli of dust formed during planet formation in protoplanetary discs.  Moreover, the Terrestrial planets of the solar system themselves are routinely envisioned to have formed from a narrow annulus of solid debris \citep{HANSEN:2009}.  By carrying out the analysis of this work, we have also lifted one of the chief limitations of the approach adopted by \citep{BAT:2018}: the assumption that the dynamics are effectively local.  While this breaks the theoretical tractability of the normal modes, it provides a self-consistent method of finding the normal modes and their frequency of evolution, as well as ways to compute angular momentum and the rate of change of angular momentum. Nevertheless, there are several important limitations to our updated model.  First, the dynamics considered in this paper occur only on the secular timescale; this model fails to capture dynamics that occur on timescales comparable to the orbital period or comparable to the lifespan of the system.  Correspondingly, the orbit-averaged nature of the dynamics discussed in this paper preclude it from describing some interesting behavior. 

Second, while this paper usefully extends the local-coupling Schr{\"o}dinger model of the secular small-inclination dynamics of a razor-thin self-gravitating particle disc, significant work remains. Most prominently, we assumed that the dynamics were generated only through purely gravitational forces, and that the disc is a particle disc.  While neglecting pressure forces is a good approximation if $\sqrt{\mathcal{G}\Sigma a}$ is much greater than the speed of sound \citep{TREMAINE:2001}, any comprehensive theory needs to account for these internal effects.  Other authors have argued that discs evolving purely through pressure forces, neglecting viscosity and self-gravity, can be described by a non-linear Schr{\"o}dinger equation.  Incorporating this formalism with the theory contained in this paper may allow us to extend this work to fluid pressure-supported discs. Indeed, such a development constitutes an interesting avenue for future work.

\section*{Acknowledgements}

W. M. is grateful to the Caltech Summer Undergraduate Research Fellowship program, during which this work was initiated. K.B. is grateful to the David and Lucile Packard Foundation and the Alfred P. Sloan Foundation for their generous support. Additionally, we would like to thank the anonymous referees for providing useful insights that have led to an improvement of the manuscript.

\section*{Data Availability}
The data underlying this article will be shared on reasonable request to the corresponding author.





\bibliographystyle{mnras}
\bibliography{tbib1}



\appendix

\section{A Direct Proof of The Conservation of the Inner Product}
\label{DIP}
In this appendix, we present a proof that the inner product $\braket{\eta}{\nu}$ defined in Equation \ref{IPE} is conserved without assuming that angular momentum ($\langle\eta|\eta\rangle$) is conserved. Earlier, we took the continuum limit of the disturbing function using a logarithmic parametrization of radial distance $\rho = \log a/a_{\rm{in}}$. Once we take the continuum limit, we can choose to reparametrize the problem in terms of another coordinate $\chi$, such that the semimajor axis of the infinitesimal disc parametrized by $\chi$ is $a(\chi)$.  In terms of $\chi$, the Hamiltonian describing the coupling becomes
\begin{equation}
\mathcal{H}\eta = \int C(a(\chi),a(\chi')(\eta(\chi)-\eta(\chi'))2\pi a(\chi')\Sigma(\chi')\frac{da}{d\chi'}d\chi',
\end{equation}
with coupling between the points $\chi$ and $\chi'$ being described by the function 
\begin{equation}
B(\chi,\chi') = 2\pi a(\chi')\Sigma(a(\chi'))\frac{da}{d\chi}(\chi')C(a(\chi),a(\chi')).
\end{equation}
Suppose that we choose $a(\chi)$ such that $B(\chi,\chi') = B(\chi',\chi)$.  Rearranging this equation, we find that this implies that 
\begin{equation}
\frac{da}{d\chi}(\chi) = \frac{a(\chi')\Sigma(a(\chi'))C(\chi,\chi')}{a(\chi)\Sigma(a(\chi))C(\chi',\chi)}\frac{da}{d\chi}(\chi').
\end{equation}
Because 
\begin{align*}
C(a,a')C(a',a'')C(a'',a) = C(a',a)C(a'',a')C(a,a''),
\end{align*} this equation provides a consistent definition of $da/d\chi$. With this parametrization, we can find a classical Hamiltonian that describes the full system:
\begin{equation}
\begin{split}
\mathcal{H}_c &= \iint B(\chi,\chi')(\eta(\chi)\eta(\chi')^*)d\chi d\chi'  \\
 &- \int\left(\int B(\chi,\chi')d\chi'\right)\eta(\chi)\eta(\chi)^*d\chi.
\end{split}
\end{equation}
Following the derivation above with this parametrization leads to a Hermitian Hamiltonian with respect to the inner product\newline  $\braket{\eta}{\nu} = \int \eta^*\nu d\chi$ \citep{STROCCHI:1966}. Incorporating formula (14) above, we have that the conserved inner product is 
\begin{equation}
\braket{\eta}{\nu} \propto \int \eta^*(a)\nu(a)\frac{a\Sigma(a)C(a_0,a)}{a_0\Sigma(a_0)C(a,a_0)da/d\chi(a_0)}da.
\end{equation}
For our choice of disc parameters, we have that $\braket{\eta}{\nu} = \int \eta^*\nu ada$ as can be directly computed from the above expression. For arbitrary surface density profile $\Sigma(a)$, the inner product derived here takes the form given by Equation \ref{IPE}. We have not assumed that $\braket{\eta}{\eta}$ is conserved; hence, we have proved that this model conserves angular momentum. 

\bsp	
\label{lastpage}
\end{document}